\documentclass[acmsmall]{acmart}

\usepackage{tabularx}
\AtBeginDocument{%
  \providecommand\BibTeX{{%
    \normalfont B\kern-0.5em{\scshape i\kern-0.25em b}\kern-0.8em\TeX}}}

\setcopyright{acmcopyright}
\copyrightyear{2021}
\acmYear{2021}
\acmDOI{doi}

\begin{document}

\title{Parlermonium: A Data-Driven UX Design Evaluation of the Parler Platform}

\author{Emma Pieroni}
%\authornote{College of Computing and Digital Media, School of Design}
 \affiliation{%
   \institution{DePaul University}
   \streetaddress{243 S Wabash Ave}
   \city{Chicago}
   \state{IL}
   \postcode{60604}
   \country{USA}
 }
 \email{epieroni@depaul.edu}

\author{Peter Jachim}
%\authornote{College of Computing and Digital Media, School of Design}
 \affiliation{%
   \institution{DePaul University}
   \streetaddress{243 S Wabash Ave}
   \city{Chicago}
   \state{IL}
   \postcode{60604}
     \country{USA} 
 }
 \email{pjachim@depaul.edu}

\author{Nathaniel Jachim}
 \affiliation{
   \institution{University of Michigan}
   \streetaddress{243 S Wabash Ave}
   \city{Dearborn}
   \state{MI}
   \postcode{60604}
     \country{USA} 
 }
 \email{njachim@umich.edu}

\author{Filipo Sharevski}
 %\authornote{College of Computing and Digital Media, School of Computing}
 \affiliation{
   \institution{DePaul University}
   \streetaddress{243 S Wabash Ave}
   \city{Chicago}
   \state{IL}
   \postcode{60604}
   \country{USA}
 }
 \email{fsharevs@cdm.depaul.edu}

\renewcommand{\shortauthors}{E. Pieroni, P. Jachim, N. Jachim, F. Sharevski}
%\renewcommand{\shortauthors}{Authors}

\begin{abstract}
This paper evaluates Parler, the controversial social media platform, from two seemingly orthogonal perspectives: user design perspective and data science. UX design researchers explore how users react to the interface/content of their social media feeds; Data science researchers analyze the misinformation flow in these feeds to detect alternative narratives and state-sponsored disinformation campaigns. We took a critical look into the intersection of these approaches to understand how Parler's interface itself is conductive to the flow of misinformation and the perception of ``free speech'' among its audience. Parler drew widespread attention leading up to and after the  2020 U.S. elections as the ``alternative'' place for free speech, as a reaction to other mainstream social media platform which actively engaged in labeling misinformation with content warnings. Because platforms like Parler are disruptive to the social media landscape, we believe the evaluation uniquely uncovers the platform's conductivity to the spread of misinformation. 
\end{abstract}

%   <concept>
%       <concept_id>10002978.10003029.10011703</concept_id>
%       <concept_desc>Security and privacy~Usability in security and privacy</concept_desc>
%       <concept_significance>500</concept_significance>
%       </concept>
%   <concept>
%       <concept_id>10002978.10002997.10003000.10011612</concept_id>
%       <concept_desc>Security and privacy~Phishing</concept_desc>
%       <concept_significance>500</concept_significance>
%       </concept>
%  </ccs2012>

% \ccsdesc[500]{Security and privacy~Usability in security and privacy}
% \ccsdesc[500]{Security and privacy~Phishing}

% \keywords{phishing, interaction-based training, Amazon Alexa, intelligent voice assistants}


\maketitle

\section{Introduction}
Since the rise of social media in the United States, there has been a fierce debate over the role, legal and societal, of media companies in regulating and moderating speech on their platforms \cite{Balkin}. With the First Amendment of the U.S. Constitution protecting freedom of speech, some view means of platform moderation as abridging that right \cite{Hooker}. This debate came into acute focus during the 2020 U.S. Election cycle, when Twitter and Facebook escalated use of warnings and deleted content in which it deemed misleading or harmful to public discourse \cite{Roth}. Seeking to escape the ``policing'' of these platforms, many fled to a relatively low-profile platform named Parler, which dedicates itself to ``free speech-driven" policies and prides itself as a ``public square" of opinions \cite{Matze}. An early hint to such a ``platform migration,'' although in the opposite direction, was noticed from alt-right communities within 4chan and Reddit toward Twitter \cite{Caulfield}. In this paper, our objective is not to analyze the debate over social media moderation nor the flow of alternative narratives between platforms, but rather, to understand how a platform's interaction with ``free speech'' can manifest itself in the user design and reflect on its audience. 

What happened inconsistently and irregularly before the 2020 U.S. election, culminated in the aftermath, when misinformation about the legitimacy of the election spread rapidly across many social media platforms \cite{cisa}. Mainstream platforms like Facebook and Twitter began adding labels to President Trump's account content regularly. The intentions and incentives of social media companies in these instances are outside the purview of this paper, but moderation like this has contributed to a majority of Americans --and in particular, Republican-leaning Americans-- who believe that social media companies censor certain political perspectives \cite{Vogels}. Unlike other social media platforms, Parler stood out, as it elected to keep content on its platform that had been otherwise removed by large Silicon Valley companies. Those wary of excessive censorship began to flock to Parler, boasting ``banned by the mainstream media'' as a badge of honor, and as more people migrated off mainstream platforms, Parler became a safe haven for \textit{all} opinions, as it became host to a wide range of conspiracy theories and fringe alternative groups. Naturally, in the quest to understand the intricacy of how trolling narratives develop on social media, it became clear we needed to investigate this previously-obscure platform further, which brings us to the topic of this paper: A data-driven UX design analysis of the Parler platform to uncover the attraction of the site as well as any intentions to target that appeal at specific groups of users.

% \subsection{Outline}

\section{Parler: Aesthetics}
% \includegraphics[]{Screenshots/UI - Home Page.jpg}
% Requesting a crisper screenshot.

\subsection{Rustic Appearance}
Parler was founded in 2018, although it didn't gain large popularity until 2020. Since the platform has been live for multiple years, to some extent, Parler's design and development teams have had ample opportunity to fix issues that are deemed problematic, and have had time to implement ambitious design features as they wish, as shown in Figure 1. This means that, to some extent, we can assume that obvious glitches have been ignored for a reason or are intentionally left alone. For example, the rounded corners on many pages appear slightly jagged. Additionally, while a dark theme is offered, many of the buttons in the dark theme do not appear to change color, forcing the users' eyes to re-adjust. The lack of polish on the site demonstrates the clear disparity between Parler, and the conventional Silicon Valley social media companies like Twitter and Facebook. Implicitly, this contrast emphasizes Parler's ``outsider'' status, appealing to users who deem themselves ``other" as well. A similar effort to stay distinctly outsider with even more crude appearance of contrasting red/orange background and ruby font (the ``Yotsuba'' style) or a similar dark theme (the ``Tomorrow'' style) could also be observed on the infamous discussion board 4chan ~\cite{Tuters}.  

\begin{figure}[h]
  \centering
  \includegraphics[width=0.5\linewidth]{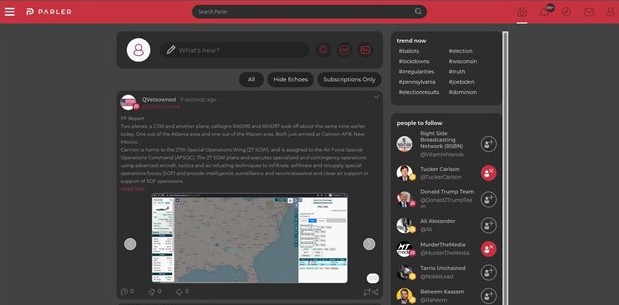}
  \caption{Parler: Home Page}
\end{figure}

% \subsubsection{Dark Theme}
% Parler offers a dark theme to help encourage users to spend longer on their application while reducing eye strain.

\subsection{Relative Anonymity}
While viewing a parley, the small font size, profile picture, and lack of color contrast pulls attention away from the source, allowing the user to focus more intently on the content of the parley, regardless of source reliability. Broadly, the potential for anonymity online allows users to open up about their true feelings, without having to take responsibility, by separating their actions from the real world and identity \cite{Nissenbaum}. By obscuring the identity of the poster of a parley, this incentives users to post with more genuine intention, without the fear of repercussions or criticism outside of Parler, while also encouraging users to avoid source-checking the information they are absorbing while on the platform. Relative anonymity is a desired feature for platforms supporting progressive dissent, for example 4chan, which enables ``ephemeral design'' where posts are deleted after a certain amount of user engagement \cite{Bernstein}. The absence of persistent markers of identity and reputation, common to social media platforms, enables Parler users, like users on 4chan, to continually demonstrate their subcultural status of ``others'' and reformulate the boundaries of their community \cite{Nissenbaum}.

\subsection{Importance of Images}
When a user clicks on a profile page to learn more or investigate further, the use of a monochromatic grey-scale between text and background provides little contrast. So instead of drawing a user's eye to text, a user is prone to focus on the images. Visual symbols play a significant role in constructing political narratives and can be used to evoke strong emotions surrounding a political event that words cannot \cite{Lilleker}. Images are often used to amplify an event, by adding drama, but can also serve as means to identify an event or cause efficiently, through communicating specific arguments \cite{Schill}. In the context of Parler, the focus on the large profile banner gives a user a unique opportunity to present a compelling argument --in a fraction of the time-- through an image, rather than relying on text to convey that narrative. Using the profile page for user \@QStorm1111, shown in Figure 2, whose background image utilizes a combination of blue lightning and large text, as well as a dark color scheme; this picture is quite striking, and to a viewer, might elicit a narrative that there is some dramatic or ``biblical" change coming to Washington. A similar utilization of images also can be observed on 4chan with the anonymous sharing of \textit{memes}, which allow otherwise complete strangers to demonstrate and negotiate in-group belonging through their vernacular fluency \cite{Nissenbaum}.
%Unlike Parler, however, these memes or provocative profile images are somewhat crudely displayed in each post with a link to the meme/image file and no relative adjustment to the textual content.

\begin{figure*}[h]
  \centering
  \includegraphics[width=0.45\linewidth]{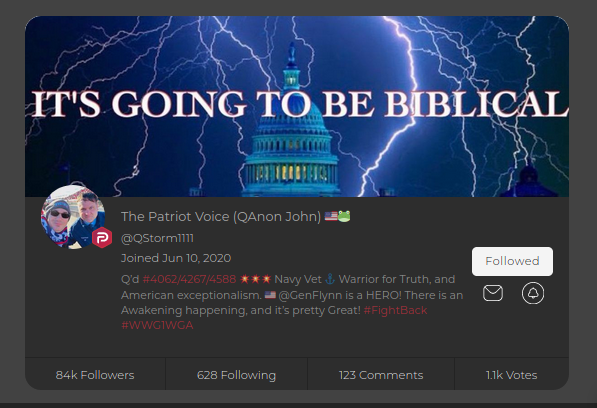}
  \caption{Parler: A Cover Photo Example}
\end{figure*}

\subsection{Color}
The Parler platform is composed of a dominantly red and grey color scheme, in which red is prone to elicit more powerful emotions than many other colors \cite{Elliot}. But the specific emotions associated with a color are dependent on the environment or context that the color is presented in; in achievement or intense situations, red is used to evoke strong emotions, as it is associated with negative connotations such as threat, danger, and aggression -- the competitive nature of social media may fit this context \cite{Elliot}. Alternatively, as in the case of 4chan (the Yotsuba style), the red color can signal the subculture status of antagonism. In both cases, one could argue that the red is a deliberate choice of the \textit{dark pattern} design approach, geared toward \cite{Gray}. 

Rather than moving a user's eye across the page by using a tertiary color to offset the red/grey color scheme, instead, users are drawn to the specific elements which are highlighted in red or any additional features that take up space on the page. Grey-on-grey text provides little contrast or visual interest when reading, and even the deep red tone appears muted on its grey background (also, red on grey is not color-blindness friendly; a condition of color blindness as \textit{protanopia} makes the red color appear black resulting in a confusing look). Instead of focusing on the procedural information or dulled text, attention is drawn to the images throughout the page, further emphasizing the importance of visuals on the site.

\section{Parler: Functionality}

\subsection{Search Function}
% This is Peter's observation, may not fit from a UX perspective, feel free to cut.
Unlike Twitter and even 4chan, where a user can search for a term or a word and view tweets/posts that have used that term or word, in Parler, a user can only search for usernames, or for specific hashtags. This has a couple of different effects on the platform. The obfuscation of search terms helps protect users from being searched for online, protecting unprotected speech. While use of certain words can make a person unemployable, this means that someone doing research cannot search to see occurrences of slurs or other objectionable parlance that could be brought to, for example, an individual's employer. This provides a slim layer of protection for individual users, allowing them to say what they like and only to make specific words ``searchable''.  Additionally, this layer of protection can help encourage people to make slanderous claims. For example, if Pfizer wants to see any misinformation about their vaccine, they cannot simply search vaccine, but have to follow specific individuals to see what they say about the vaccine. This gives individuals like Alex Jones freedom to make claims that the COVID-19 vaccines contain HIV, as shown in Figure 3, without risk of litigation, when Pfizer was the only vaccine with FDA emergency use approval. 

% ~\cite{fda_pfizer, fda_moderna}. 

\begin{figure*}[h]
  \centering
  \includegraphics[width=0.45\linewidth]{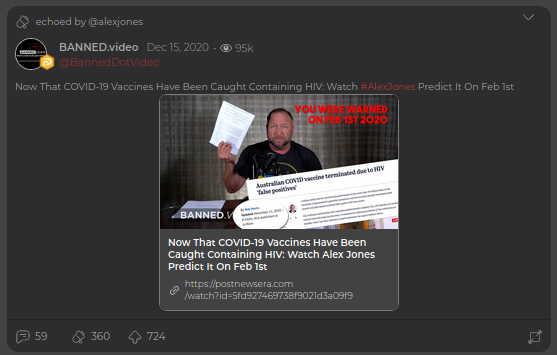}
  \caption{A Parler Post Claiming the COVID-19 Vaccine Contains HIV}
\end{figure*}

The limited search functionality also affects the user's freedom to investigate alternate perspectives regarding specific events. Unlike Twitter, where a user can search all tweets in the aftermath of an event to see a variety of different perspectives, on Parler, users can only search for parleys that have specific hashtags. The end result, is that users have much less freedom to see a variety of viewpoints on events, but are rather confined to comfortable and familiar hashtags. The ineffectiveness is psychologically convenient, as instinctively, people tend toward homophily, or the tendency for similar individuals to associate with each other more often than with individuals viewed as ``different'' \cite{Grevet}. Somewhat contradictory to Parler's emphasis on ``free speech'', the inability to find additional or alternative perspectives, leaves users predestined to focus on one comfortable narrative, cognitively segregated from other perspectives, as well as inevitably leaving Parler to fail its own expectations of creating a ``public square''-style exchange of ideas.

\subsection{People to Follow}
There is a box on the right side that tracks a long list of accounts that a user can follow as shown in Figure 4. A close inspection of these accounts reveals that they belong to ``others'' or the alternative, self-proclaimed campus of free speech defenders \cite{Vogels}. On the surface, this resembles a common functionality across other social media sites. LinkedIn, for example, suggests ``People you may know,'' and Twitter offers a list of ``Who to Follow.'' The stark difference between these features is that once a user is added from the ``People you may know'' or ``Who to Follow'' lists, that user is removed from the list, and new recommended users are added. In contrast to Twitter or LinkedIn, Parler's functionality serves as a checklist, where the checklist shows people who a user already follows, indicated with a red check-mark, and continues to re-recommend the users who are not followed yet. This can be interpreted as a progress-indicator, implicitly suggesting these are the people a user \textit{should} follow, in order to have a \textit{complete} feed. Illustrated by the popular ``gotta catch 'em all'' phrase from Pokemon, the \textit{Zeigarnik Effect} explains that incomplete tasks are more potent and memorable than completed ones \cite{Zeigarnik}, leaving a user more prone to continue playing Pokemon or, translated in the context of Parler, continue following profiles until the given task is completed. 

\begin{figure}[h]
  \centering
  \includegraphics[width=0.25\linewidth]{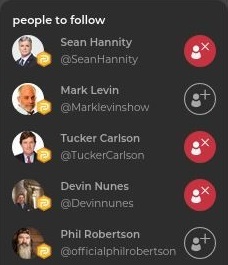}
  \caption{Parler: People to Follow Frame}
\end{figure}

As shown in Figure 4, Parler's ``People to Follow" window consists of almost entirely of users flaunting a yellow Parler badge, identifying them as ``verified influencers'' or members of the Parler community with a large following. By recommending these large influencers rather than regular accounts, Parler amplifies the voices of its pundits rather than its users, contradicting its CEO's, John Matze, perspective that the objective of Parler is to be a ``public square for promoting discourse'' \cite{Matze}. Rather than other social media platforms that opt for the terms ``likes'', ``shares'', or ``retweets'', Parler has coined ``echoes'' to quantify interaction on a post, but the interface's focus on influencers takes the term to a new level, as it turns the platform into an echo-chamber for its largest users \cite{Grevet}. For example, an early analysis of the Parler structure found that about 40\% of the typical users have more than a single follower, whereas about 40\% of gold badge users have more than 10,000 followers (the verified influences fall somewhere in the middle) \cite{Aliapoulios}. %For individual users that are not identified as influencers, there is still a chance to gain a badge, by submitting a copy of an official piece of identification, such as a Driver's License, as well as a photograph of themselves. After verifying the ID and encrypting an copy to store internally, Parler issues the user a red profile badge, deeming the account a ``Real User'' and differentiating it from potential troll and bot accounts \cite{Matze}. 

\section{Data Analysis of Parler's Content Flows}
Given the isolated nature of the Parler platform, to build a feed to collect data from, we elected to follow a variety of different accounts (including @TheProudBoys, @InfoWars, @WarRoomPandemic, @KAGDonaldTrump, @qanonymous, amongst others). Some of these accounts stand out as conservative thought-leaders and were selected via Parler's ``People to Follow" feature, and another portion were found via searching selected terms and hashtags including \#StopTheSteal, \#AmericaFirst, \#MAGA, and ``qanon.'' After expanding our network on Parler, we compiled a dataset between September 2020 and January 21 2021 (before Parler was taken offline). In total, the dataset contains more than 200,000 words, representing opinions presented by 67 Parler users, and approximately 10,000 parleys. To comply with Parler's terms and conditions, we manually copied and pasted data from accounts we followed, and pasted them into an Excel file.

\subsection{Correspondence Analysis}
Our first approach to analyzing trending discussions on Parler was through the correspondence analysis shown in Figure 5. A correspondence analysis takes the eigenvectors of a contingency table based on the $\chi^2$ statistic. To create the contingency table, we compared the top nouns and verbs used in parleys. We parsed parleys into their parts of speech, and cleaned to just reference their lemma. The first dimension, which accounts for 46.78\% of the variability in the data, indicates a forceful narrative around votes and the Capitol, and a more hopeful and reflective narrative around Twitter and the President. Higher values in the first dimension include scores for the nouns ``Capitol,'' ``America,'' ``People,'' and ``Vote,'' while the verbs used are mostly forceful verbs, like ``make,'' ``take,'' and ``say.'' Lower values include hopeful and reflective verbs, like ``will,'' ``know,'' ``want,'' and ``come,'' as well as the nouns ``Twitter,'' ``President,'' and ``election.'' 

\begin{figure*}[h]
  \centering
  \includegraphics[width=0.5\linewidth]{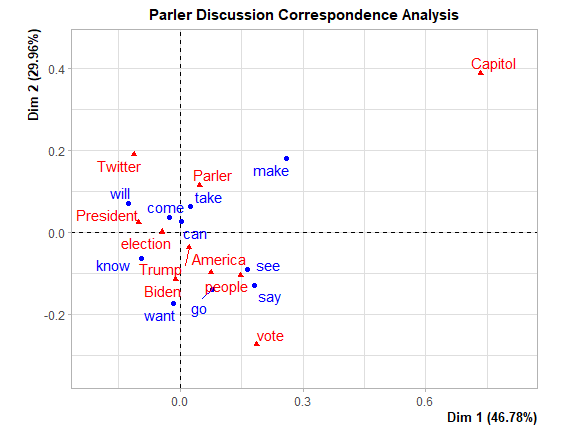}
  \caption{Parler Content Flow Correspondence Analysis}
\end{figure*}

The second dimension, which accounts for 29.96\% of the variability in the data, suggesting that ``Twitter,'' ``Parler,'' and the ``Capitol'' are all part of a movement with stronger, forceful verbs, while verbs used to describe votes, people, America, and Biden are more optimistic (hopeful that Biden will be somehow ousted). High values for the second dimension start with the noun ``Capitol'' and continues to the noun ``vote'', with higher values in the second dimension describing ``Twitter'' and ``Parler.'' The verbs associated with those nouns are forceful/challenging verbs, like ``make,'' ``take,'' `will,'' and ``come.'' Lower values in the second dimension seem to indicate a more positive, aspirational interpretation of these nouns (it is, however, worth noting that the positive association with Biden may, for example, have to do with how Parler users ``want'' Biden to lose, ``go'' away, ``say'' Trump won the election, etc.).

% The correspondence analysis accounts for 76.74\% of the total variability in the contingency table that is not captured in the expected value.

%  using the SpacyR \cite{spacyr} library To perform this analysis, we used the FactoMineR R library \cite{FactoMineR}.

\subsection{Regression Analysis}
Our second analysis was performed using a regression analysis to see the extent we could model the interactions between different words. This gives some idea of the extent that all of the topics and verbs discussed were related to one another. To model this information, we built a second-order linear regression model that estimated the log of the total number of times each noun-verb pair was used based just on the number of nouns and the number of verbs in the contingency table created above. Here is the model used:
\vspace{-0.1em}
$$ln(tpc)=-3.723+1.063e^{-3}*vc+1.873e^{-6}*nc-1.242e^{-7}*vc^2+9.991*log(nc)+\epsilon$$

where $tpc$ is the total number of times that a noun/verb pair appeared, $vc$ is the number of times that verb was used, $nc$ is the number of times each noun was used, and $\epsilon$ represents the residual errors that are not accounted for by the model. The model accounted for 91\% of the variability in the total number of times each pair of words appeared, and had an adjusted $R^2$ of 90.7\%. In practical terms, this means, in addition to the specific deviations discussed above, all verbs and nouns were used together quite frequently as well. So, while ``Twitter'' and ``Capitol'' appeared separately in the correspondence analysis above, they were discussed in similar contexts as well. A close inspection of the content suggest the most dominant context may be Twitter's ban on Trump following the insurgencies on the Capitol building.

\subsection{Linguistic Analysis}
We performed a linguistic analysis of the collected dataset as well a subset of the parleys containing the words ``Twitter'' and ``Capitol'', shown in Table 1 \cite{Pennebaker}. The higher ``analytical thinking'' score indicates that the overall parleys follow logical and considerate thinking patterns. This is not surprising, given Parler's somewhat ``refined'' reputation compared to other sub-cultural platforms, e.g. 4chan. The Twitter/Capitol parleys show an even higher score on the analytical thinking, indicating that they are mostly direct ``here-and-now'' expressions. The high ``clout'' score indicates the high level of confidence that Parler users display through their writing. This high confidence stems from the beliefs and convictions to defend ``free speech'' online. An even higher ``clout'' score of the Twitter/Capitol parleys suggest even higher confidence, characteristic for the ``verified influencers'' on the platform. The significantly low ``authenticity'' score indicates that Parler users do not post authentic content, even in the case of the Twitter/Capitol parleys. Rather, they repetitively regurgitate similar narratives promulgated by the ``verified influencers.'' Finally, the ``emotional tone'' score (below 50) indicates a negative and impolite style of writing overall, characteristic for conspiracy theories, fake news, and misinformation. The emotional tone of the Twitter/Capitol parleys, with the highest negative difference in relative score, is considerably negative and rude. This demonstrates the antagonistic sentiment about the Twitter censorship decision. 

\begin{table}[h]
\renewcommand{\arraystretch}{1.1}
\renewcommand{\tabcolsep}{2mm}
\caption{LIWC scores of the Parler Dataset.}
\label{table2}
\centering
\begin{tabular}{|l|c|c|c|c|}
\hline
 \textbf{Dataset} & \textbf{Analytical Thinking} & \textbf{Clout} & \textbf{Authenticity} & \textbf{Emotional Tone} \\
\hline
\textit{\textbf{Parler Overall}} & 87.56 & 75.91 & 16.64 &	35.67 \\\hline
\textit{\textbf{Twitter/Capitol}} & 91.72 & 83.83 & 16.17 & 20.96 \\\hline
\end{tabular}
\end{table}

\section{Conclusion}
In this paper, we sought to investigate how UX design choices aid the proliferation of alternative narratives on the Parler platform. First, we analyzed the design and functionality features of the platform itself, and then, collected and analyzed a dataset to understand how content flows on the platform. Our findings suggest that Parler is especially conducive to dissemination of misinformation through a variety of factors including heightened user anonymity; the prioritization of images rather than parley source/text content; emphasis on platform influencers; the absence of counter-argumentation. The dominant red/grey color scheme of Parler's user interface elicits antagonistic connotations associated with the platform, a notion supported by our correspondence and linguistic analyses, which suggested overwhelming negative parley sentiment, particularly around the Twitter's decision for censorship following the failed coup on the Capitol. Designed for a ``public square'' exchange of ideas, Parler fails in delivering that ideal, and instead, it rewards and amplifies its largest users. Taking all these things into account, Parler's commitment to ``free speech'' seems largely one-sided, encouraging the spread of unverified claims, rumors, and misinformation. 

\bibliographystyle{ACM-Reference-Format}
\bibliography{parlermonium}

\end{document}